\newif\ifeprint\eprinttrue

\ifeprint

\documentclass
[rmp,twocolumn,amsfonts,amssymb,amsmath,showkeys,letterpaper,
raggedbottom,floatfix]{revtex4}
\makeatletter\def\raggedcolumn@skip{\vskip\z@\@plus.0001fil\relax}\makeatother
\def\zbreak{\notag\\&\quad}
\let\xbreak=\zbreak
\def\ybreak{}

\else

\documentclass[final,twocolumn,natbib]{svjour3}
\journalname{J. Geod.}
\usepackage{amsmath}
\usepackage{amssymb}
\usepackage{mathptmx}
\def\zbreak{\notag\\&\quad}
\def\xbreak{}
\let\ybreak=\zbreak

\fi

\date{February 8, 2010; revised \today}

\def\TITLE{Transverse Mercator with an accuracy of a few nanometers}
\usepackage{ifpdf}
\usepackage{times}
\usepackage{path}
\usepackage[bookmarks=false]{hyperref}
\ifpdf\else\usepackage{breakurl}\fi
\hypersetup{pdftitle={\TITLE},pdfauthor={C. F. F. Karney}}
\usepackage{graphicx}

\newcount\minutes \newcount\hours
\hours=\time \divide\hours by 60 \multiply\hours by 60
\minutes=\time \advance\minutes by -\hours \divide\hours by 60

\def\twodigit#1{\ifnum#1<10 0\fi\the#1}

\defcitealias{dlmf10}{{\sc dlmf}}
\defcitealias{krueger12}{Kr\"uger}
\def\dlmf#1#2{\href{http://dlmf.nist.gov/#2}{#1}}
\def\eprint#1{E-print \href{http://arxiv.org/abs/#1}{\rm arXiv:\penalty100 #1}}
\def\doi#1{doi:\discretionary{}{}{}\href
{http://dx.doi.org/#1}{{\let\tt=\rm\path|#1|}}}
\def\tfrac#1/#2 {{\textstyle\frac{#1}{#2}}}
\newcommand{\gd}{\mathop{\mathrm{gd}}\nolimits}
\newcommand{\sech}{\mathop{\mathrm{sech}}\nolimits}
\newcommand{\sn}{\mathop{\mathrm{sn}}\nolimits}
\newcommand{\am}{\mathop{\mathrm{am}}\nolimits}
\def\abs#1{\left|#1\right|}
\def\sqrta#1{\sqrt{\vphantom{\sin^2u}\smash{#1}}}

\begin{document}

\ifeprint
\noindent\mbox{\begin{minipage}[b]{\textwidth}
\begin{flushright}
Link \url{http://geographiclib.sf.net/tm.html}\par\vspace{0.5ex}
\eprint{1002.1417}\par
\vspace{2ex}
\end{flushright}
\end{minipage}
\hspace{-\textwidth}}
\fi

\title{\TITLE}
\ifeprint
\author{\href{http://charles.karney.info}{Charles F. F. Karney}}
\email{charles.karney@sri.com}
\affiliation{\href{http://www.sri.com}{SRI International},
201 Washington Rd, Princeton, NJ 08543-5300}
\else
\author{Charles F. F. Karney}
\institute{C. F. F. Karney \at
\href{http://www.sarnoff.com}{SRI International},
201 Washington Rd, Princeton, NJ 08543-5300\\\email{charles.karney@sri.com}
}
\fi

\ifeprint\else\maketitle\fi

\begin{abstract}

Implementations of two algorithms for the transverse Mercator projection
are described; these achieve accuracies close to machine precision.  One
is based on the exact equations of Thompson and Lee and the other uses
an extension of Kr\"uger's series for the mapping to higher order.
The exact method provides an accuracy of $9\,\mathrm{nm}$ over the
entire ellipsoid, while the errors in the series method are less than
$5\,\mathrm{nm}$ within $3900\,\mathrm{km}$ of the central meridian.  In
each case, the meridian convergence and scale are also computed with
similar accuracy.  The speed of the series method is competitive with
other less accurate algorithms and the exact method is about 5 times
slower.

\ifeprint
\else
\fi
\end{abstract}

\ifeprint\maketitle\fi

\section{Introduction}\label{intro}

The transverse Mercator or Gauss--Kr\"uger projection is a conformal
mapping of the earth ellipsoid where a central meridian is mapped into a
straight line at constant scale.  Because it cannot be expressed in
terms of elementary functions, the mapping is usually computed by means
of a truncated series \citep{krueger12,thomas52}.  The resulting mapping
approximates the true mapping only within a region centered on the
central meridian.

Transverse Mercator is one of the commonest projections used for
large-scale maps (it is used for the grid systems of several countries
and is the basis of the universal transverse Mercator (UTM)
system \citep[Chap.~2]{utmups}).  For the WGS84 ellipsoid, the variation
of the scale is $1.25\%$ within $1000\,\mathrm{km}$ of the central
meridian; it is therefore desirable to find algorithms for the mapping
which are accurate to machine precision over at least this area.  In
this paper, I describe the implementation of two such algorithms, one
based on the exact equations given by \citet{lee76} and the other
extending the series given by \citet{krueger12} to higher order.  Both
implementations compute the forward and reverse mappings and also return
the meridian convergence and scale.  These implementations are included
in GeographicLib \citep{geographiclib17}.

Scores of other authors have presented methods for computing this
mapping over the past century.  In particular, \citet{dozier80} provided
an implementation of Lee's exact method, and \citet{engsager07} give
Kr\"uger's series to 7th order.  The distinguishing aspects of this work
are the reduction of the overall numerical errors (truncation and
round-off) to close to the precision limit of the computer and the
concrete bounds I place on these errors.

Because floating-point numbers have a finite
spacing \citep[\S\dlmf{3.1(i)}{3.1.i}]{dlmf10}, the limiting
accuracy of any implementation is about $M/2^p$ where $M =
10\,000\,\mathrm{km}$ is the length of the quarter meridian of the earth
and $p$ is the number of bits in the fraction of the floating-point
number system.  This gives an error limit of $0.5\,\mathrm m$ for $p =
24$ (single precision or float), $1\,\mathrm{nm}$ for $p = 53$ (double
precision or double), and $0.5\,\mathrm{pm}$ for $p = 64$ (extended
precision or long double).  (Here, I use SI prefixes: $1\,\mathrm{nm} =
10^{-9}\,\mathrm m$, $1\,\mathrm{pm} = 10^{-12}\,\mathrm m$.)  Typically
$p=24$ is too inaccurate to be useful and I don't consider this further
in this paper.  My standard working precision is double and the
resulting accuracy, if it can be achieved, would satisfy most needs.
However, I also use extended precision as one of the tools to verify the
accuracy of the double precision implementations.

Formulas for mappings can contain expressions which are numerically
ill-conditioned causing precision to be lost.  This loss of precision is
of little consequence if the truncation errors are of the same order.
However, in attempting to minimize the numerical errors, I needed an
accurate means of quantifying the truncation and round-off errors.  To
this end, I constructed a large test set of projected points which were
computed with an accuracy of 80 decimal digits.  This allowed me to
eliminate many sources of round-off error.  The resulting accuracies are
about 4--8 times the limiting value (equivalent to a loss of only 2--3
bits of precision) and this applies to both double and extended
precisions.

In Sect.~\ref{krug}, I review the series method given
by \citet{krueger12} modifying it to minimize the round-off errors.  I
turn next, Sect.~\ref{exact}, to the formulation of the exact transverse
Mercator projection by \citet{lee76} which I use to construct the
high-precision test set; I also describe its implementation using double
precision and I quantify the round-off errors.  I extend Kr\"uger's
series to 8th order (see Sect.~\ref{series}) and give the truncation
error for the series as a function of truncation level and distance from
the central meridian.  Finally, in Sect.~\ref{prop}, I discuss some of
the properties of the exact mapping far from the central meridian.

\section{Kr\"uger's series}\label{krug}

I summarize here the method developed by \citet[\S\S5--8]{krueger12},
simplifying it and adapting it for optimal implementation on a computer.
The method is also briefly described by \citet[\S5.1.6]{bugayevskiy95}.
The method entails mapping the ellipsoid to the conformal sphere and for
this reason I begin by describing the spherical transverse Mercator
projection.

Consider a sphere and a point on that sphere of latitude $\phi'$ and
longitude relative to the central meridian of $\lambda$.  (I use primes
on variables, e.g., $\phi'$, where necessary, to distinguish them from
their ellipsoidal counterparts.)  The isometric latitude is given by
\begin{equation}\label{psi0}
\psi' = \gd^{-1} \phi',
\end{equation}
where
\[
\gd x = \int_0^x \sech t\, dt = \tan^{-1}\sinh x = \sin^{-1}\tanh x
\]
is the Gudermannian function given
by \citet[\S\dlmf{4.23(viii)}{4.23.viii}]{dlmf10} (henceforth
referred to as \citetalias{dlmf10}, the Digital Library of Mathematical
Functions) and
\[
\gd^{-1} x = \int_0^x \sec t\, dt = \sinh^{-1}\tan x = \tanh^{-1}\sin x
\]
is its inverse.  The standard (equatorial) Mercator projection maps the
sphere onto the plane $(\lambda, \psi')$.  When working with conformal
mappings it is often useful to represent coordinates with complex
numbers where the real part represents the northing and the imaginary
part the easting (the phase, or argument, of the complex number gives a
bearing measured {\it clockwise}).  In this representation the Mercator
projection (a conformal mapping) is given by
\[
\chi = \psi' + i\lambda.
\]
Any analytic function of $\chi$ also represents a conformal mapping
(except where its derivative vanishes); its derivative gives the change
in the meridian convergence and scale for the mapping.  In
particular \citep[Eq.~(12.3)]{lee76},
\begin{equation}\label{spher-tm}
\zeta' = \gd \chi = \gd (\psi' + i\lambda)
\end{equation}
gives the transverse Mercator projection of the sphere.  This is easy to
confirm by evaluating the mapping for $\lambda = 0$; this gives $\zeta'
= \phi'$, i.e., the central meridian is mapped to a straight line at
constant scale (the defining property of the mapping).

I consider now an ellipsoid of revolution with equatorial radius $a$,
polar semi-axis $b$, flattening $f = (a-b)/a$, eccentricity $e
= \sqrt{f(2-f)}$, and third flattening $n = (a-b)/\allowbreak(a+b) =
f/(2-f)$.  For a point with latitude $\phi$ and longitude $\lambda$,
the isometric latitude is given by \citep[\S117]{lambert72}
\[
\psi =
\log\tan\biggl(\frac\pi4 + \frac{\phi}2\biggr)
-\frac12 e
\log \biggl(\frac{1 + e \sin\phi}{1 - e \sin\phi}\biggr),
\]
Using the identities \citetalias[Eqs.~\dlmf{(4.23.42)}{4.23.E42}
and \dlmf{(4.37.24)}{4.37.E24}]{dlmf10}, this relation may also be
written as
\begin{equation}\label{psi1}
\psi = \gd^{-1}\phi - e\tanh^{-1}(e \sin\phi).
\end{equation}
As in the case of the sphere, $\chi = \psi + i\lambda$ defines the
Mercator projection.  Equating the isometric latitude for the sphere
with that for the ellipsoid, $\psi' = \psi$,
defines a relation
\begin{equation}\label{conformal0}
\phi' = \gd\bigl(\gd^{-1}\phi - e\tanh^{-1}(e \sin\phi)\bigr),
\end{equation}
which maps a point on the ellipsoid with latitude $\phi$ conformally to
a point on the sphere with latitude $\phi'$.  In this context, $\phi'$
is called the ``conformal latitude'' and the sphere is referred to as
the ``conformal sphere.''  The transformation to $\zeta'$,
Eq.~(\ref{spher-tm}), where $\psi$ is given by Eq.~(\ref{psi1}), defines
a conformal mapping of the ellipsoid to a plane
in which the central meridian is
mapped to a straight line with a scale
which is nearly constant (the variation is $O(f)$).  I call this the
``spherical transverse Mercator projection'' as it is the simplest
generalization of the spherical projection to the ellipsoid.  The
graticule for this mapping is shown in
Fig.~\ref{graticule}(a).  \citet[\S8]{krueger12} now ``rectifies''
this mapping by applying a near-identity transformation to $\zeta'$ to
make the scale constant which yields the Gauss--Kr\"uger mapping $\zeta
= \xi + i \eta$ with
\begin{equation}\label{forward}
\zeta = \zeta' + \sum_{j=1}^\infty \alpha_j \sin 2j\zeta',
\end{equation}
where $\alpha_j$ is real (this form for the transformation is derived in
Sect.~\ref{series}).  Similarly the transformation from $\zeta'$ to
$\zeta$ can be written as
\begin{equation}\label{reverse}
\zeta' = \zeta - \sum_{j=1}^\infty \beta_j \sin 2j\zeta,
\end{equation}
where $\beta_j$ is real.  Kr\"uger's expressions for $\alpha_j$ and
$\beta_j$ are given below, Eqs.~(\ref{alpha}) and (\ref{beta}), and we
outline their derivation in Sect.~\ref{series}.

First, I address the computation of $\zeta'$ given $\phi$ and $\lambda$
with special emphasis on maintaining numerical accuracy.  Following
Kr\"uger, I write $\zeta' = \xi' + i \eta'$ and give separate equations
for $\xi'$ and $\eta'$.  However, in order to maintain accuracy near
$\phi = \pm \frac12 \pi$, I use $\tau = \tan\phi$ and $\tau'
= \tan\phi'$ and eliminate $\phi'$ and $\psi$ from the relations.  An
expression for $\tau'$ is found by taking the tangent of
Eq.~(\ref{conformal0}) and using the addition rule for the hyperbolic
sine to give
\begin{equation}\label{conformal}
\tau' = \tau \sqrt{1 + \sigma^2} - \sigma \sqrt{1 + \tau^2},
\end{equation}
where
\begin{align}
\tau &= \tan \phi,\label{taueq}\\
\label{sigma}
\sigma &= \sinh\bigl(e \tanh^{-1}(e \tau/\sqrt{1 + \tau^2}) \bigr).
\end{align}
Eliminating $\phi'$ from the expressions for $\xi'$ and
$\eta'$ \citep[Eq.~(8.36)]{krueger12} yields
\begin{equation}\label{fzetap}
\begin{split}
\xi' &= \tan^{-1}(\tau' / \cos\lambda),\\
\eta' &= \sinh^{-1}\bigl(\sin\lambda \big/
         \sqrt{\tau'^2 + \cos^2\lambda}\bigr).
\end{split}
\end{equation}
Splitting Eq.~(\ref{forward}) into real and imaginary parts gives
\begin{equation}\label{fzeta}
\begin{split}
\xi &= \xi' + \sum_{j=1}^\infty \alpha_j \sin 2j \xi' \cosh 2j\eta',\\
\eta &= \eta' + \sum_{j=1}^\infty \alpha_j \cos 2j \xi' \sinh 2j\eta',
\end{split}
\end{equation}
where \citep[Eq.~(8.41)]{krueger12}
\begin{equation}\label{alpha}
\begin{split}
\alpha_1 &= \frac12 n
           - \frac23 n^2
           + \frac5{16} n^3
           + \frac{41}{180} n^4
           + \cdots,\\
\alpha_2 &= \frac{13}{48} n^2
           - \frac35 n^3
           + \frac{557}{1440} n^4
           + \cdots,\\
\alpha_3 &= \frac{61}{240} n^3
           - \frac{103}{140} n^4
           + \cdots,\\
\alpha_4 &= \frac{49561}{161280} n^4
           + \cdots.
\end{split}
\end{equation}
Finally, $\xi$ and $\eta$ are scaled to give the transverse
Mercator easting $x$ and northing $y$,
\begin{equation}\label{eastnorthf}
x = k_0A\eta,\quad y = k_0A\xi,
\end{equation}
where $k_0$ is the scale on the central meridian, $2\pi A$ is the
circumference of a meridian, and \citep[Eq.~(5.5)]{krueger12}
\begin{equation}\label{rad}
A = \frac a{1+n} \biggl(1 + \frac14  n^2 + \frac1{64} n^4 + \cdots \biggr).
\end{equation}
Typically $k_0$ is chosen to be slightly less than $1$ to minimize the
deviation of the scale from unity in some region around the central
meridian.

Converting from transverse Mercator to geographic coordinates entails
reversing these steps.  Equations (\ref{eastnorthf}) give
\begin{equation}\label{eastnorthr}
\eta = x/(k_0A),\quad \xi = y/(k_0A).
\end{equation}
\citet[\S7]{krueger12} writes $\zeta'$ in
terms of $\zeta$ by inverting Eq.~(\ref{fzeta}) to give
\begin{equation}\label{rzetap}
\begin{split}
\xi' &= \xi - \sum_{j=1}^\infty \beta_j \sin 2j \xi \cosh 2j\eta,\\
\eta' &= \eta - \sum_{j=1}^\infty \beta_j \cos 2j \xi \sinh 2j\eta,
\end{split}
\end{equation}
where \citep[Eq.~(7.26*)]{krueger12}
\begin{equation}\label{beta}
\begin{split}
\beta_1 &= \frac12 n
           - \frac23 n^2
           + \frac{37}{96} n^3
           - \frac1{360} n^4
           + \cdots,\\
\beta_2 &= \frac1{48} n^2
           + \frac1{15} n^3
           - \frac{437}{1440} n^4
           + \cdots,\\
\beta_3 &= \frac{17}{480} n^3
           - \frac{37}{840} n^4
           + \cdots,\\
\beta_4 &= \frac{4397}{161280} n^4
           + \cdots.
\end{split}
\end{equation}
Inverting Eq.~(\ref{fzetap}) gives \citep[Eq.~(7.25)]{krueger12}
\begin{equation}\label{rpsi}
\begin{split}
\tau' &= \sin\xi'\big/
        \sqrta{\sinh^2\eta' + \cos^2\xi'},\\
\lambda &= \tan^{-1}(\sinh\eta'/\cos\xi').
\end{split}
\end{equation}
Equation (\ref{conformal}) may be inverted by Newton's method,
\begin{align}
\tau_i &=
\begin{cases}
\tau', & \text{for $i = 0$},\\
\tau_{i-1} + \delta\tau_{i-1}, & \text{otherwise},
\end{cases}\\
\tau'_i &= \tau_i \sqrta{1 + \sigma_i^2} - \sigma_i \sqrta{1 + \tau_i^2},\\
\delta\tau_i &= \frac
{\tau' - \tau'_i}{\sqrta{1 + \tau'^2_i}}
\frac{1 + (1 - e^2)\tau_i^2}{(1 - e^2) \sqrta{1 + \tau_i^2}}.
\end{align}
This usually converges to round-off after two iterations, i.e., $\tau
= \tau_2$, which gives
\begin{equation}\label{phieq}
\phi = \tan^{-1}\tau.
\end{equation}

The meridian convergence and scale can be found during the forward
mapping by differentiating Eq.~(\ref{forward}) and writing
\[
p' - i q' = \frac{d\zeta}{d\zeta'},
\]
or
\begin{equation}\label{fpq}
\begin{split}
p' &= 1 + \sum_{j=1}^\infty 2j \alpha_j \cos 2j \xi' \cosh 2j\eta',\\
q' &= \sum_{j=1}^\infty 2j \alpha_j \sin 2j \xi' \sinh 2j\eta'.
\end{split}
\end{equation}
Then the meridian convergence (the bearing of grid north, the $y$ axis,
measured clockwise from true north) is given by $\gamma = \gamma'
+ \gamma''$, where \citep[Eqs.~(8.44--45)]{krueger12}
\begin{equation}\label{fconv}
\begin{split}
\gamma' &= \tan^{-1}\bigl( (\tau'/\sqrt{1 + \tau'^2}) \tan\lambda \bigr),\\
\gamma'' &= \tan^{-1}(q'/p').
\end{split}
\end{equation}
The scale is given by $k = k_0 k' k''$, where
\citep[Eq.~(8.47)]{krueger12}
\begin{equation}\label{fscale}
\begin{split}
k' &= \sqrta{1 - e^2 \sin^2\phi} \sqrt{1 + \tau^2} \big/
      \sqrt{\tau'^2 + \cos^2\lambda},\\
k'' &= \frac Aa \sqrt{p'^2+q'^2}.
\end{split}
\end{equation}
Here $\gamma'$ and $k'$ give the convergence and scale for the spherical
transverse Mercator projection, while $\gamma''$ and $k''$ give the
corrections due to Eqs.~(\ref{forward}) and (\ref{eastnorthf}).

To determine the convergence and scale during the reverse mapping,
differentiate Eq.~(\ref{reverse}) and write
\[
p + i q = \frac{d\zeta'}{d\zeta} = \frac1{p' - i q'},
\]
or
\begin{equation}\label{rpq}
\begin{split}
p &= 1 - \sum_{j=1}^\infty 2j \beta_j\cos 2j \xi \cosh 2j\eta,\\
q &= \sum_{j=1}^\infty 2j \beta_j \sin 2j \xi \sinh 2j\eta.
\end{split}
\end{equation}
The convergence is given by $\gamma = \gamma' + \gamma''$,
where \citep[Eqs.~(7.31--31*)]{krueger12}
\begin{equation}\label{rconv}
\begin{split}
\gamma' &= \tan^{-1}(\tan\xi' \tanh\eta'),\\
\gamma'' &= \tan^{-1}(q/p).
\end{split}
\end{equation}
The scale is given by $k = k_0 k' k''$, where
\citep[Eq.~(7.33)]{krueger12}
\begin{equation}\label{rscale}
\begin{split}
k' &= \sqrta{1 - e^2 \sin^2\phi} \sqrt{1 + \tau^2}
      \sqrta{\sinh^2\eta' + \cos^2\xi'},\\
k'' &= \frac Aa \frac1{\sqrt{p^2+q^2}}.
\end{split}
\end{equation}

In summary, Kr\"uger's methods for the forward and reverse mappings are
given by the numbered Eqs.~(\ref{conformal})--(\ref{rad}) and
Eqs.~(\ref{rad})--(\ref{phieq}), respectively.  The scale and meridian
convergence are similarly given by the Eqs.~(\ref{fpq})--(\ref{fscale})
during the forward mapping and Eqs.~(\ref{rpq})--(\ref{rscale}) during the
reverse mapping.

\citetalias{krueger12} truncates the series at order $n^4$, as shown here.
This results in very small errors, considering that Kr\"uger published
his paper in 1912.  The maximum of the errors for the forward and
reverse mappings (both expressed as true distances) is $0.31\,\mu\mathrm
m$ within $1000\,\mathrm{km}$ of the central meridian and is
$1\,\mathrm{mm}$ within $6000\,\mathrm{km}$ of the central meridian.
The truncated mapping is exactly conformal; however Eqs.~(\ref{forward})
and (\ref{reverse}) are not inverses of one another if the sums are
truncated.  It is, of course, possible to construct an exact inverse of
the truncation of Eq.~(\ref{forward}), e.g., by solving it using
Newton's method.  However, in practice, it is better merely to retain
enough terms in the sum so that the truncation error is less than the
round-off error.

In numerically implementing this method, the terms $A$, $\alpha_j$, and
$\beta_j$, Eqs.~(\ref{rad}), (\ref{alpha}), and (\ref{beta}), need only
be computed once for a given ellipsoid and, for accuracy and speed,
should be evaluated in Horner
form \citepalias[\S\dlmf{1.11(i)}{1.11.i}]{dlmf10}; for example,
$\alpha_1$ is evaluated to order $n^4$ as
\[
\alpha_1 =\bigl(\tfrac 1/2 + \bigl(-\tfrac 2/3
+(\tfrac 5/16 + \tfrac 41/180 n)n\bigr)n\bigr)n.
\]
Furthermore the trigonometric series, Eqs.~(\ref{fzeta}),
(\ref{rzetap}), (\ref{fpq}), and (\ref{rpq}), can be evaluated
using \citet{clenshaw55}
summation \citepalias[\S\dlmf{3.11(ii)}{3.11.ii}]{dlmf10} which
minimizes the number of evaluations of trigonometric and hyperbolic
functions.  Thus Eqs.~(\ref{fzeta}) and (\ref{fpq}) may be summed to
order $J$ with
\begin{align*}
c_{J+1} &= c_{J+2} = 0, \\
c_j &= 2 c_{j+1} \cos 2(\xi' + i\eta') - c_{j+2} + \alpha_j, \\
\xi + i\eta  &= \xi' + i\eta' + c_1 \sin 2(\xi' + i\eta'),
\end{align*}
and
\begin{align*}
d_{J+1} &= d_{J+2} = 0, \\
d_j &= 2 d_{j+1} \cos 2(\xi' + i\eta') - d_{j+2} + 2 j \alpha_j, \\
p' + i q' &= 1 - d_2 + d_1 \cos 2(\xi' + i\eta'),
\end{align*}
separated into real and imaginary parts and with the recursion relations
for $c_j$ and $d_j$ evaluated for $J \ge j > 0$.  The summations of
Eqs.~(\ref{rzetap}) and (\ref{rpq}) are handled in a similar fashion.

My introduction to Kr\"uger's expansion was a report by the Finnish
Geodetic Institute \citep{jhs154}.  The method described here follows
this report with a few changes to improve the numerical accuracy: (a)~I
use more stable formulas for converting from geographic to the spherical
transverse Mercator coordinates; (b)~I solve for the geographic latitude
by Newton's method instead of by iteration; and (c)~I use Kr\"uger's
method for determining the convergence and scale instead of less
accurate expansions in the longitude.

In contrast to the series given here, the formulas given by
\citetalias{krueger12} in a later section of his paper, \S14, involve
an expansion in the longitude difference instead of the flattening.
This expansion forms the basis of the approximate transverse Mercator
formulas presented by \citet[pp.~2--6]{thomas52} and in the report on
UTM \citep[Chap.~2]{utmups} and are used in \citet{geotrans}.  For
computing UTM coordinates, the errors are less than
$1\,\mathrm{mm}$.  Unfortunately, the truncated series does not define
an exact conformal mapping.  In addition, in some applications, use of
these series may lead to unacceptably large errors.  For example,
consider mapping Greenland with transverse Mercator with a central
meridian of $42^\circ\,\mathrm W$.  The landmass of Greenland lies
within $750\,\mathrm{km}$ of this central meridian and the maximum
variation in the scale of transverse Mercator is only $0.7\%$---in other
words, the transverse Mercator projection is ideal for this application.
The error in computing transverse Mercator with Kr\"uger's 4th order
series is (as we have seen) less than $1\,\mu\mathrm m$.  However the
maximum error using Thomas' series (as implemented in Geotrans, version
3.0) is over $1\,\mathrm{km}$.

\section{Exact mapping}\label{exact}

The definition of the transverse Mercator projection given at the
beginning of Sect.~\ref{intro} serves to specify the mapping completely.
(There are two minor qualifications to this statement: the central
scale, the origin, and the orientation of the central meridian need to
be specified; in addition, the mapping becomes multi-valued very far
from the central meridian as detailed in Sect.~\ref{prop}.)  Provided
that the series in Eqs.~(\ref{forward}), (\ref{reverse}), and
(\ref{rad}) are convergent, the Kr\"uger series method converges to
the exact Gauss--Kr\"uger projection and the truncated series are a
useful basis for numerical approximations to the mapping.

There is no problem with the convergence of expression for $A$,
Eq.~(\ref{rad}).  This can be written in closed form as
\[
A = \frac{2a}\pi E(e)
  = \frac{2a}\pi E\bigl(4n/(1+n)^2\bigr),
\]
where $E(k)$ is the complete elliptic integral of the second kind with
modulus $k$ \citepalias[Eq.~\dlmf{(19.2.8)}{19.2.E8}]{dlmf10}, which
may be expanded in a series
using \citetalias[Eq.~\dlmf{(19.5.2)}{19.5.E2}]{dlmf10} to give
\begin{equation}\label{radext}
A = \frac a{1+n} \bigl(1 + \tfrac1/4 n^2
                           + \tfrac1/64 n^4
                           + \tfrac1/256 n^6
                           + \tfrac25/16384 n^8 + \cdots \bigr).
\end{equation}
This series converges for $\abs n < 1$ and, for small $n$, the relative
error in truncating the series is given by the first dropped term.

The convergence of Eqs.~(\ref{forward}) and (\ref{reverse}) is more
complicated because the sine terms in the summands become large for
large $\eta$ or $\eta'$.  Indeed, the transverse Mercator projection has
a singularity in its second derivative at $\phi = 0^\circ$ and $\lambda
= \pm\lambda_0$ where $\lambda_0 = (1 - e)90^\circ$ beyond which the
series will diverge; these points are branch points of the
mapping \citep[\S5.7]{whittaker27} and the properties of the mapping in
their vicinity are explored in Sect.~\ref{prop}.  In order to determine
the error in the truncated series, I implement the formulas for the {\it
exact} mapping as given by \citet[\S\S54--55]{lee76} who credits
E. H. Thompson (1945) for their development.  A referee has pointed out
to me that a similar formulation was independently provided
by \citet{ludwig43}.  Here I give only a brief description of Lee's
method, referring the reader to the documentation and source code for
GeographicLib for more details \citep{geographiclib17}.

The exact mapping is expressed in terms of an intermediate
mapping, the Thompson projection,
denoted by $w = u + i v$ with \citep[Eqs.~(54.5) and
(55.5)]{lee76}
\begin{align}
\chi &= \tanh^{-1}\sn w - e \tanh^{-1}( e \sn w ), \label{chiw}\displaybreak[0]\\
\zeta &= \frac\pi{2 E(e)} \bigl( E(e) - \mathcal E(K(e)-w, e) \bigr),
\label{zetaw}
\end{align}
where $\sn u$ is one of the Jacobi elliptic functions with modulus
$e$ \citepalias[\S\dlmf{22.2}{22.2}]{dlmf10}, $K(k)$ is the complete
elliptic integral of the first kind with modulus
$k$ \citepalias[Eq.~\dlmf{(19.2.8)}{19.2.E8}]{dlmf10}, and $\mathcal
E(x, k)$ is Jacobi's epsilon
function \citepalias[Eq.~\dlmf{(22.16.20)}{22.16.E20}]{dlmf10}.

When implementing these equations, I follow Lee and break the formulas
in terms of their real and imaginary parts.  This enables the algorithm
to be implemented with real arithmetic which allows the expressions to
be optimized to minimize the round-off error.  The necessary formulas
for Eqs.~(\ref{chiw}) and (\ref{zetaw}) are given
by \citet[Eqs.~(54.17) and (55.4)]{lee76}.

The computation of the forward (resp.~reverse) mapping requires the
inversion of Eq.~(\ref{chiw}) (resp.\ Eq.~(\ref{zetaw})).  I perform
these inversions using Newton's method in the complex plane.
The needed
derivative of $\chi$ is given by \citet[Eq.~(54.21)]{lee76} and
$d\zeta/dw$ is given by \citet[Eq.~(55.9)]{lee76} which may be
split into real and imaginary parts
with \citetalias[Eq.~\dlmf{(22.8.3)}{22.8.E3},
\S\dlmf{22.6(iv)}{22.6.iv}]{dlmf10}.
The starting guesses for Newton's method are obtained by finding
approximate solutions using one of three methods: (a)~by using the limit
$e \rightarrow 0$, (b)~by expanding about the branch point on the
equator (the bottom right corner of Fig.~\ref{extend}(c)), or (c)~by
expanding about the singularity at the south pole (the top right corner
of Fig.~\ref{extend}(c)).  (The latter two methods require a knowledge
of the properties of the mapping far from the central meridian; see
Sect.~\ref{prop} for more information.)  The most time-consuming task in
this implementation was optimizing the choice of starting point to
ensure that the method converges in a few iterations.  I refer the
reader to the code for details. I also compute the meridian convergence
and scale using \citet[Eqs.~(55.12--13)]{lee76}.

In order to reduce the round-off errors, I needed to identify terms in
the formulas with the potential for a loss of precision and apply
identities for the Jacobi elliptic
functions \citepalias[Eq.~\dlmf{(22.2.10)}{22.2.E10},
\S\dlmf{22.6(i)}{22.6.i}]{dlmf10} to recast the
formulas into equivalent ones with better numerical properties.  I use
the procedure $\mathit{sncndn}$ given by \citet{bulirsch65} for the
elliptic functions and algorithms $R_F$, $R_D$, and $R_G$
of \citet{carlson95} for the elliptic integrals; these algorithms can
yield results to arbitrary precision.

I provide two implementations of the exact mapping: (a)~a C++ version
using standard floating-point arithmetic and (b)~an implementation
in \citet{maxima}.  The latter implementation makes use of Maxima's
``bigfloat'' package which permits the calculation to be carried out to
an arbitrary precision.  This was used to construct a large test set for
the mapping which served to benchmark the C++ implementation.  This set
includes randomly distributed points together with additional points
chosen close to the pole and other possibly problematic points and
lines.  The mapping is computed to an accuracy of 80 decimal digits and
the results are rounded to the nearest $0.1\,\mathrm{pm}$.  Both the C++
and Maxima implementations of the exact mapping and the test data are
provided with GeographicLib \citep{geographiclib17}.

The C++ implementation was checked by computing the maximum of the error
in the forward mapping expressed as a true distance (i.e., dividing
the error in the mapped space by the scale of the mapping) and the
error in the reverse mapping (again expressed as a true distance).
When implemented using double (resp.~extended) precision, the maximum
round-off error is $\delta_r = 9\,\mathrm{nm}$ (resp.~$5\,\mathrm{pm}$)
over the whole ellipsoid (using the WGS84 parameters, $a =
6\,378\,137\,\mathrm m$ and $f = 1/298.257\,223\,563$).  These are
consistent with $\delta_r \approx M/2^{p-3}$ indicating that the error
is only about 8 times the limiting round-off error given in
Sect.~\ref{intro}.  The truncation error $\delta_t$,
defined in Sect.~\ref{series}, is zero for this
method.

Using the double precision implementation, the errors in the meridian
convergence and scale at a particular point are bounded by
\begin{align*}
\delta\gamma_r &< \frac1{2^{p-3}} \biggl(1 +
\frac M{s_p} + 1.5 \sqrt[3]{\frac M{s_b}} \biggr)
\frac{180^\circ}\pi,\\
\frac{\delta k_r}k &< \frac1{2^{p-3}} \biggl(1 +
1.5 \sqrt[3]{\frac M{s_b}} \biggr),
\end{align*}
where $s_p$ and $s_b$ are the geodesic distances from the point to the
closest pole and closest branch point, respectively.  These bounds were
found empirically; however the form of the expressions is determined by
the nature of the singularities in the mapping.  The term involving
$s_p$ arises because small errors in the position close to the pole may
cause large changes in the convergence.  Similarly the terms involving
$s_b$ appear because of the singularity in the second derivative of the
mapping which causes the convergence and scale to vary rapidly near the
branch point.

\begin{figure}[t]
\begin{center}
    \vspace{0mm}\mbox{\hspace{-1mm}\includegraphics
      [width=85mm,angle=0]{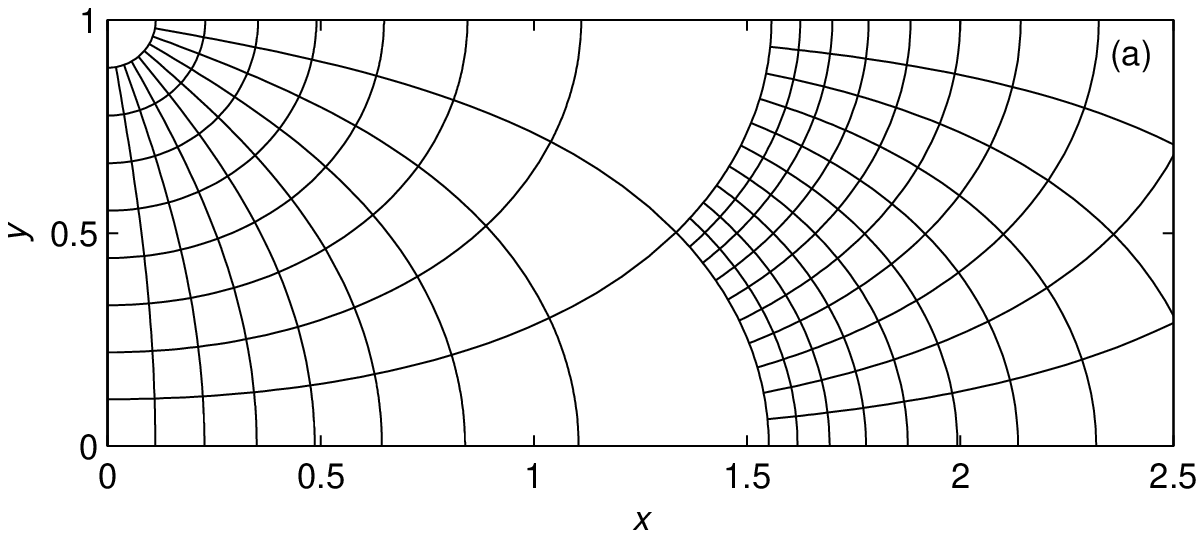}}%
\vspace{-4mm}
\par\vspace{5mm}
    \vspace{0mm}\mbox{\hspace{-1mm}\includegraphics
      [width=85mm,angle=0]{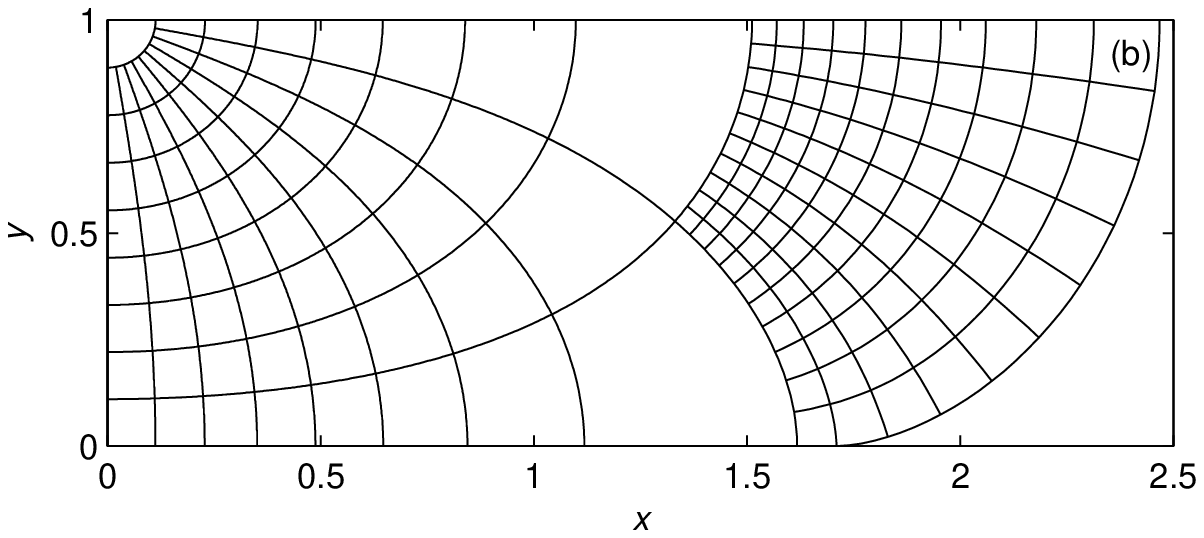}}%
\vspace{-4mm}
\par\vspace{5mm}
    \vspace{0mm}\mbox{\hspace{-1mm}\includegraphics
      [width=85mm,angle=0]{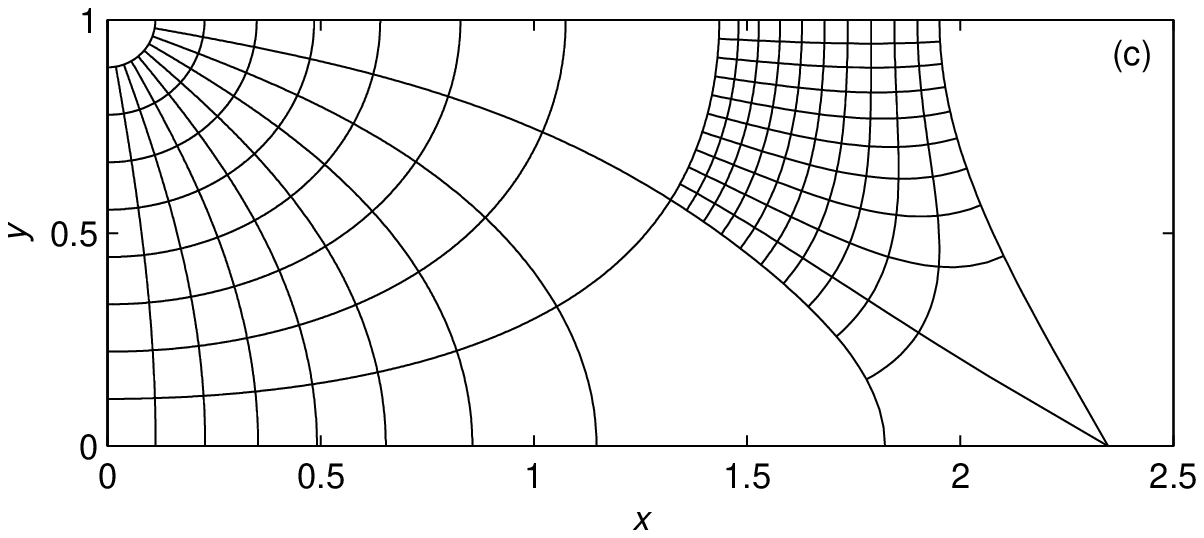}}%
\vspace{-2mm}
\end{center}
\caption
{Graticules for the (a)~spherical transverse Mercator,
(b)~Gauss--Kr\"uger, and (c)~Thompson projections.  Here $x$ and $y$ are
the easting and northing for the mappings.  The eccentricity is $e
= \frac1{10}$ ($f \approx 1/199.5$) and the mappings have been scaled so
that the distance from the equator to the north pole is unity.  Thus
$\lambda = 0^\circ$ maps to the line $x = 0$ and $\lambda = 90^\circ$
maps to the line $y = 1$.  The graticule is shown at multiples of
$10^\circ$ with $1^\circ$ lines added in $80^\circ < \lambda < 90^\circ$
and $0 < \phi < 10^\circ$.}
\label{graticule}
\end{figure}
The differences between my implementation and that of \citet{dozier80}
are as follows.  (a)~Dozier's starting guesses for Newton's method are
based only on the limit $e \rightarrow 0$.  Newton's method then fails
to converge in the neighborhood of the branch point (where ellipsoidal
effects become large).  In contrast, I use different methods for
computing the starting points in different regions which enables
Newton's method to converge everywhere.  (b)~I modified several of the
equations to improve the numerical accuracy.  Without this, Dozier loses
about half the precision in some regions.  (c)~I use published
algorithms to evaluate the special
functions \citep{bulirsch65,carlson95}.  (d)~I compute the meridian
convergence and scale.  (e)~Lastly, I provide an arbitrary precision
implementation (in Maxima) to allow the errors in the C++ implementation
to be measured accurately.

Figures \ref{graticule}(b) and (c) show the graticule for the
Gauss--Kr\"uger and Thompson projections.  One eighth of the ellipsoid is
shown in these figures, $0^\circ \le \phi \le 90^\circ$ and
$0^\circ \le \lambda \le 90^\circ$, and (unlike the spherical transverse
Mercator projection, Fig.~\ref{graticule}(a)) this maps to a finite area
with the Gauss--Kr\"uger and Thompson projections.  To obtain the
graticule for the entire ellipsoid, reflect these figures in $x = 0$, $y
= 0$, and $y = 1$.  The eccentricity for these figures is $e
= \frac1{10}$ and, in this case, the equator runs along $y = 0$ until
the branch point at $\lambda = \lambda_0 = 81^\circ$ and then heads for
$y=1$; the point $\phi = 0^\circ$, $\lambda = 90^\circ$ maps to finite
points on $y = 1$.  The Thompson mapping is not conformal at the branch
point (where there's a kink in the equator), because $d\chi/dw$ vanishes
there.  Similar figures are given by \citet[Figs.~43--46]{lee76}.

\section{Extending Kr\"uger's series}\label{series}

There are several ways that the series for $A$, $\alpha_j$ and $\beta_j$
can be generated; here, I adopt an approach which is close to that used
by \citet[\S5]{krueger12}.  (Alternative methods are to expand
Eqs.~(\ref{chiw}) and (\ref{zetaw}) or to use the polar stereographic
projection instead of the Mercator projection as the starting
point \citep{wallis92}.)  In the limit $\eta = \eta' = 0$ (i.e., on the
central meridian, $\lambda = 0$), the quantities $\zeta$ and $\zeta'$
become the rectifying and conformal latitudes respectively.  (The
conformal latitude $\phi'$ was introduced in Sect.~\ref{krug}; the
rectifying latitude is linearly proportional to the distance along a
meridian measured from the equator.)  The transformation between $\zeta$
and $\zeta'$ is thus given by the relation between the rectifying and
conformal latitudes extended to the complex plane.  Thus $\zeta$ is the
meridian distance scaled to $\pi/2$,
\begin{equation}\label{zetaphi}
\zeta(\Phi) = \frac\pi{2E(e)}
\int_0^{\Phi} \frac{1-e^2}{(1-e^2\sin^2\phi)^{3/2}}\, d\phi,
\end{equation}
where $\Phi$ is the normal geographic latitude extended to the complex
plane.  The integral here can be expressed in terms of elliptic
integrals as $E(e) - E(\Theta, e)$, where $E(\phi, k)$ is the incomplete
elliptic integral of the second kind with argument $\phi$ and modulus
$k$ \citepalias[Eq.~\dlmf{(19.2.5)}{19.2.E5}]{dlmf10}, and $\Theta
= \cot^{-1}\bigl((1-f)\tan\Phi\bigr)$ is the parametric co-latitude.
Similarly, $\zeta'$ is merely Eq.~(\ref{conformal0}) extended to the
complex plane.
\begin{equation}\label{zetapphi}
\zeta'(\Phi) = \gd \bigl( \gd^{-1}\Phi - e \tanh^{-1}( e \sin\Phi ) \bigr).
\end{equation}
The quantities $\Phi$ and $\Theta$ are related to the Thompson projection
variable $w$ by
\begin{equation}\label{phiw}
\begin{split}
\Phi &= \am w,\quad \Theta = \am(K(e) - w),\\
w &= F(\Phi, e) = K(e) - F(\Theta, e),
\end{split}
\end{equation}
where $\am w$ is Jacobi's amplitude
function \citepalias[\S\dlmf{22.16(i)}{22.16.i}]{dlmf10} with modulus
$e$ and $F(\phi, k)$ is the incomplete elliptic integral of the first
kind with argument $\phi$ and modulus
$k$ \citepalias[Eq.~\dlmf{(19.2.4)}{19.2.E4}]{dlmf10}.  Substituting
Eq.~(\ref{phiw}) into Eqs.~(\ref{zetapphi}) and (\ref{zetaphi}) and
using Eq.~(\ref{spher-tm})
and \citetalias[Eq.~\dlmf{(22.16.31)}{22.16.E31}]{dlmf10} gives
Eqs.~(\ref{chiw}) and (\ref{zetaw}); this establishes the equivalence of
Eqs.~(\ref{zetaphi}) and (\ref{zetapphi}) with the formulation
of \citet{lee76}.  These equations are used by \citet{stuifbergen09} as
the basis for an exact numerical method for the transverse Mercator
projection.  This is similar to (but rather simpler than) the method
of \citet{dozier80}.

The functions $\zeta(\Phi)$ and $\zeta'(\Phi)$ are analytic and so
define conformal transformations.  They can be expanded as a Taylor
series in $e^2$, or equivalently in $n$; I use the method
of \citet[\S16]{lagrange70} \citep[\S7.32]{whittaker27} to invert these
series to give the inverse functions $\Phi(\zeta)$ and $\Phi(\zeta')$.
For example, if
\[
\zeta(\Phi) = \Phi + g(\Phi),
\]
where $g(\Phi) = O(n)$, then the inverse function is
\[
\Phi(\zeta) = \zeta + h(\zeta),
\]
where
\[
h(\zeta) = \sum_{j = 1}^\infty
\frac{(-1)^j}{j!}
\left.
\frac{d^{j-1}g(\Phi)^j}{d\Phi^{j-1}}
\right|_{\Phi=\zeta}.
\]
Now compose $\zeta\bigl(\Phi(\zeta')\bigr)$ and
$\zeta'\bigl(\Phi(\zeta)\bigr)$ to provide the required series,
Eqs.~(\ref{forward}) and (\ref{reverse}).  These manipulations were
carried out using the algebraic tools provided by \citet{maxima}.
Little effort was expended to optimize this calculation since it only
needs to be carried out once!  (The expansion to order $n^8$ takes
about 15 seconds.)  At 8th order, the series for $A$ is given by
Eq.~(\ref{radext}) and the series for $\alpha_j$, Eq.~(\ref{alpha}), and
$\beta_j$, Eq.~(\ref{beta}), become
\begin{align}\label{alphaext}
\alpha_1 &= \tfrac 1/2 n
          - \tfrac 2/3 n^2
          + \tfrac 5/16 n^3
          + \tfrac 41/180 n^4
          - \tfrac 127/288 n^5
          + \tfrac 7891/37800 n^6\zbreak
          + \tfrac 72161/387072 n^7
          - \tfrac 18975107/50803200 n^8
          + \cdots,\displaybreak[0]\notag\\
\alpha_2 &= \tfrac 13/48 n^2
          - \tfrac 3/5 n^3
          + \tfrac 557/1440 n^4
          + \tfrac 281/630 n^5
          - \tfrac 1983433/1935360 n^6\xbreak
          + \tfrac 13769/28800 n^7\ybreak
          + \tfrac 148003883/174182400 n^8
          + \cdots,\displaybreak[0]\notag\\
\alpha_3 &= \tfrac 61/240 n^3
          - \tfrac 103/140 n^4
          + \tfrac 15061/26880 n^5
          + \tfrac 167603/181440 n^6
          - \tfrac 67102379/29030400 n^7\zbreak
          + \tfrac 79682431/79833600 n^8
          + \cdots,\displaybreak[0]\notag\\
\alpha_4 &= \tfrac 49561/161280 n^4
          - \tfrac 179/168 n^5
          + \tfrac 6601661/7257600 n^6
          + \tfrac 97445/49896 n^7\zbreak
          - \tfrac 40176129013/7664025600 n^8
          + \cdots,\displaybreak[0]\notag\\
\alpha_5 &= \tfrac 34729/80640 n^5
          - \tfrac 3418889/1995840 n^6
          + \tfrac 14644087/9123840 n^7\xbreak
          + \tfrac 2605413599/622702080 n^8
          + \cdots,\displaybreak[0]\notag\\
\alpha_6 &= \tfrac 212378941/319334400 n^6
          - \tfrac 30705481/10378368 n^7
          + \tfrac 175214326799/58118860800 n^8
          + \cdots,\displaybreak[0]\notag\\
\alpha_7 &= \tfrac 1522256789/1383782400 n^7
          - \tfrac 16759934899/3113510400 n^8
          + \cdots,\displaybreak[0]\notag\\
\alpha_8 &= \tfrac 1424729850961/743921418240 n^8
          + \cdots,
\end{align}
and
\begin{align}\label{betaext}
\beta_1  &= \tfrac 1/2 n
          - \tfrac 2/3 n^2
          + \tfrac 37/96 n^3
          - \tfrac 1/360 n^4
          - \tfrac 81/512 n^5
          + \tfrac 96199/604800 n^6\zbreak
          - \tfrac 5406467/38707200 n^7
          + \tfrac 7944359/67737600 n^8
          + \cdots,\displaybreak[0]\notag\\
\beta_2  &= \tfrac 1/48 n^2
          + \tfrac 1/15 n^3
          - \tfrac 437/1440 n^4
          + \tfrac 46/105 n^5
          - \tfrac 1118711/3870720 n^6\xbreak
          + \tfrac 51841/1209600 n^7\ybreak
          + \tfrac 24749483/348364800 n^8
          + \cdots,\displaybreak[0]\notag\\
\beta_3  &= \tfrac 17/480 n^3
          - \tfrac 37/840 n^4
          - \tfrac 209/4480 n^5
          + \tfrac 5569/90720 n^6
          + \tfrac 9261899/58060800 n^7\zbreak
          - \tfrac 6457463/17740800 n^8
          + \cdots,\displaybreak[0]\notag\\
\beta_4  &= \tfrac 4397/161280 n^4
          - \tfrac 11/504 n^5
          - \tfrac 830251/7257600 n^6
          + \tfrac 466511/2494800 n^7\zbreak
          + \tfrac 324154477/7664025600 n^8
          + \cdots,\displaybreak[0]\notag\\
\beta_5  &= \tfrac 4583/161280 n^5
          - \tfrac 108847/3991680 n^6
          - \tfrac 8005831/63866880 n^7\xbreak
          + \tfrac 22894433/124540416 n^8
          + \cdots,\displaybreak[0]\notag\\
\beta_6  &= \tfrac 20648693/638668800 n^6
          - \tfrac 16363163/518918400 n^7\xbreak
          - \tfrac 2204645983/12915302400 n^8
          + \cdots,\displaybreak[0]\notag\\
\beta_7  &= \tfrac 219941297/5535129600 n^7
          - \tfrac 497323811/12454041600 n^8
          + \cdots,\displaybreak[0]\notag\\
\beta_8  &= \tfrac 191773887257/3719607091200 n^8
          + \cdots.
\end{align}
GeographicLib \citep{geographiclib17} includes the Maxima code for
carrying out these expansions and the results of expanding the series
much further, to order $n^{30}$.

Equations (\ref{radext}), (\ref{alphaext}), and (\ref{betaext}) allow
the Kr\"uger method to be implemented to any order up to $n^8$.  In
order to determine which order to use in a given application, it is
useful to distinguish the truncation error (the difference between the
series evaluated exactly and the exact mapping) from the round-off
error (the difference between the series evaluated at finite precision
and the series evaluated exactly).

\begin{figure}[t]
\begin{center}
    \vspace{2mm}\mbox{\hspace{-1mm}\includegraphics
      [width=84mm,angle=0]{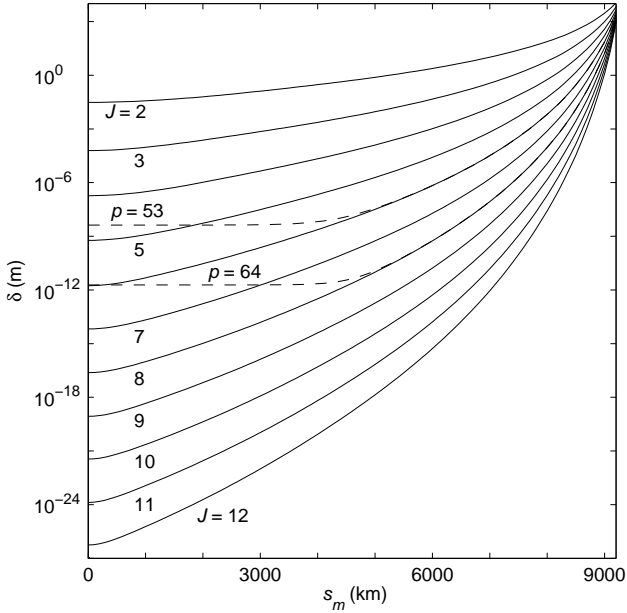}}%
\vspace{-2.5mm}
\end{center}
\caption
{Truncation and round-off errors, $\delta_t$ and $\delta_r$, in the
Kr\"uger series for the transverse Mercator projection as a function of
the distance from the central meridian $s_m$.  The series is truncated
at various orders $J$ from $2$ to $12$.  The solid lines show the
$\delta_t$.  Also shown in dashed lines are the combined round-off and
truncation errors, $\delta_t + \delta_r$ when the algorithm is
implemented with floating-point numbers with $p$ bits of fraction.  For
$p = 53$ (double) and $p = 64$ (extended), the truncation levels are set
to $J = 6$ and $J = 8$ respectively.  The WGS84 ellipsoid is used.}
\label{trunc}
\end{figure}

The truncation error was determined using Maxima's bigfloats with a
precision of 80 decimal digits.  With Kr\"uger's series the error is
principally a function of distance from the meridian (which is mainly a
function of $x$) and depends only weakly on $y$.  Defining $s_m$ as the
geodesic distance from the central meridian, I measure $\delta_t$ the
maximum of the forward and reverse truncation errors (both expressed as
true distances) over all points with a given $s_m$.  In
Fig.~\ref{trunc}, I plot $\delta_t$ as a function of $s_m$, for
truncations at various orders $J$ (the smallest terms retained are
$n^J$).  The errors rise monotonically with the distance from the
central meridian.  The branch point at $\phi = 0^\circ$ and $\lambda =
\lambda_0 \approx 82.636^\circ$ is at about $s_m =
9200\,\mathrm{km}$.  At this point the truncation error stops decreasing
with increasing order indicating a lack of convergence in the series.
(See Sect.~\ref{prop} for a proof.)  From a practical standpoint, the
convergence is too slow to be useful for $s_m \gtrsim
8000\,\mathrm{km}$.  The truncation errors in the meridian convergence
$\delta\gamma_t$ and scale $\delta k_t$ are well approximately by
\begin{align*}
\delta\gamma_t &= 2J \sec(s_m/a) \frac{\delta_t}a \frac{180^\circ}\pi,
\displaybreak[0]\\
\frac{\delta k_t}k &= 2J \sec(s_m/a) \frac{\delta_t}a.
\end{align*}
Here, the factor $2J$ arises from the differentiation performed to give
Eqs.~(\ref{fpq}) and (\ref{rpq}) and the term $\sec(s_m/a)$ is the scale
(in the spherical limit) necessary to convert the errors in position to
errors in $\zeta$ or $\zeta'$.

Figure \ref{trunc} epitomizes the advantages of Kr\"uger's over Thomas'
series.  The equivalent figure for truncation error for the latter
series would use the {\it longitude} relative the central meridian for
the abscissa instead of the {\it distance} from the central meridian.
At high latitudes, the longitude difference becomes large even for
modest distances from the central meridian; this explains the large
errors in the results from Geotrans in the Greenland example at the end
of Sect.~\ref{krug}.

Round-off errors need to be considered also when implementing the method
with floating-point arithmetic.  Evaluating the mapping using the
formulas given in Sect.~\ref{krug} adds $4.2\,\mathrm{nm}$ or
$1.9\,\mathrm{pm}$ to the truncation error depending on the precision
(see the dashed lines in Fig.~\ref{trunc}).  These round-off errors may
be expressed as $\delta_r \approx M/2^{p-2}$; i.e., they are about half
of those for the exact algorithm, because the series method involves
fewer operations.  Thus with double (resp.~extended) precision and the
series truncated at $J = 6$ (resp.~$J=8$) the overall error is less than
$5\,\mathrm{nm}$ (resp.~$2\,\mathrm{pm}$) provided that $s_m <
3900\,\mathrm{km}$ (resp.~$4200\,\mathrm{km}$).  The C++ implementation
of the mapping based on the extended Kr\"uger series (taken to order
$n^8$) is included in GeographicLib \citep{geographiclib17}.

At greater distances from the central meridian, truncation errors
become large; thus the 6th order series has an error of about
$1\,\mathrm{mm}$ at $s_m = 7600\,\mathrm{km}$ (see Fig.~\ref{trunc}).
The truncation error can be decreased by increasing the number of terms
retained in the series.  However, if high accuracy is required for
$s_m \gtrsim 4000\,\mathrm{km}$, it's probably safer to use the exact
algorithm whose error is less than $9\,\mathrm{nm}$ for the whole
spheroid.

The round-off errors in the meridian convergence and scale are bounded
by
\[
\delta\gamma_r < \frac1{2^{p-3}} \biggl(1 + 0.5 \frac M{s_p} \biggr)
\frac{180^\circ}\pi,\quad\frac{\delta k_r}k < \frac1{2^{p-3}};
\]
these should be added to $\delta\gamma_t$ and $\delta k_t$.  These
expressions have the same form as those for the exact algorithm, except
that the terms involving $s_b$ are omitted because the truncation
error dominates near the branch point.

Previously, \citet{engsager07} extended Kr\"u\-ger's series to 7th
order.  However they give a less rigorous estimate of the error.  In
addition, they give separate series for the meridian convergence and
scale, while I advocate following Kr\"uger's simpler prescription of
differentiating the same series used to carry out the mapping.  They
also give a series expansion for the transformation between latitude and
conformal latitude which leads to a somewhat faster code (see
Sect.~\ref{conclusion}); however, I prefer the simplicity of evaluating
and inverting Eq.~(\ref{conformal}) directly.

\section{Properties far from the central meridian}\label{prop}

In this section, I explore the behavior of the mapping in the vicinity
of the branch point at $\phi = 0^\circ$ and $\lambda = \lambda_0$.
First it should be remarked that this is far from the central meridian
and that the scale there is $k_0/e$; so this is not in the domain where
the mapping is very useful.  Nevertheless, it is beneficial to have a
complete understanding of a mapping; for example, this was necessary in
making the exact algorithm robust.  Here I describe the mapping in terms
of the Thompson mapping,
$w$ \citep[\S\S54--55]{lee76}.  \citet{koenig51} offer a complementary
picture based on the complex latitude $\Phi$.

The transverse Mercator projection is defined by its properties on the
central meridian and the condition of conformality.  The mapping is
defined by analytically continuing \citep[Chap.~5]{whittaker27} the
mapping away from the central meridian.  The process continues until the
branch point is encountered (where the condition of analyticity fails).
The branch point corresponds to $\chi = \chi_0 = i\lambda_0$, $\zeta
= \zeta_0 = i(K'-E')\pi/(2E)$, and $w = w_0 = iK'$, where $E = E(e)$,
$E' = E(\sqrt{1-e^2})$, $K = K(e)$, $K' = K(\sqrt{1-e^2})$, and $K(k)$
is the complete elliptic integral of the first kind with modulus
$k$ \citepalias[Eq.~\dlmf{(19.2.8)}{19.2.E8}]{dlmf10}.  The lowest
order terms in the expansions of $\chi$ and $\zeta$ about the branch
point are
\begin{align*}
\tilde\chi &= -\tfrac1/3 e(1-e^2)
\bigl[ \tilde w^3 + \tfrac1/10 (1+e^2) \tilde w^5 + \cdots \bigr],
\displaybreak[0]\\
\tilde\zeta &= -\tfrac1/3 (1-e^2)
\bigl[ \tilde w^3 + \tfrac1/5 (2-e^2) \tilde w^5 + \cdots \bigr]
\pi/(2E),
\end{align*}
where $\tilde\chi = \chi - \chi_0$, $\tilde\zeta = \zeta - \zeta_0$,
and $\tilde w = w - w_0$.  Eliminating $w$ from these equations gives
\begin{equation}
\tilde\zeta = \frac\pi{2E} \biggl[\frac{\tilde\chi}e
+ \frac{i\sqrt[3]{1-e^2}}{10}
\biggl( \frac{3i\tilde\chi}e \biggr)^{5/3}
+ O(\tilde\chi^{7/3})\biggr].\label{bp}
\end{equation}
The value of $\zeta$ will depend on how the $\frac53$ power is taken.
Picking the complex phase of $i\tilde\chi$ in the interval $(-\pi, \pi]$
gives the principal value.  Equivalently, the value of $\zeta(\chi)$ can
be made single valued by placing ``cuts'' on the equator in the
longitude ranges $(1 - e)90^\circ \le \abs\lambda \le (1 + e)90^\circ$
which act as impassable barriers during the process of analytic
continuation.  This represents the ``standard'' convention for mapping a
geographic position to the Gauss--Kr\"uger projection since the sign of
the northing matches the sign of the latitude (with the equator mapping
to non-negative northings).  This convention corresponds to
Fig.~\ref{graticule}(b) (after suitable reflections to cover the
ellipsoid), to \citet[Fig.~46]{lee76},
to \citet[Fig.~55(b)]{koenig51}, and to \citet[p.~214]{ludwig43}.

From the form of the mapping near the branch point, it is clear that the
Kr\"uger series does not converge for $\phi = 0$ and $\lambda
> \lambda_0$ because, from Eq.~(\ref{bp}), $\zeta$ is complex under
these conditions but all the terms in the series, Eq.~(\ref{forward}),
are pure imaginary.  Delineating the precise boundary for convergence of
the series and its inverse, Eq.~(\ref{reverse}), requires an analysis of
the problem for complex $e$ and is beyond the scope of this work.

\begin{figure}[t]
\begin{center}
    \vspace{0mm}\mbox{\hspace{-1mm}\includegraphics
      [width=85mm,angle=0]{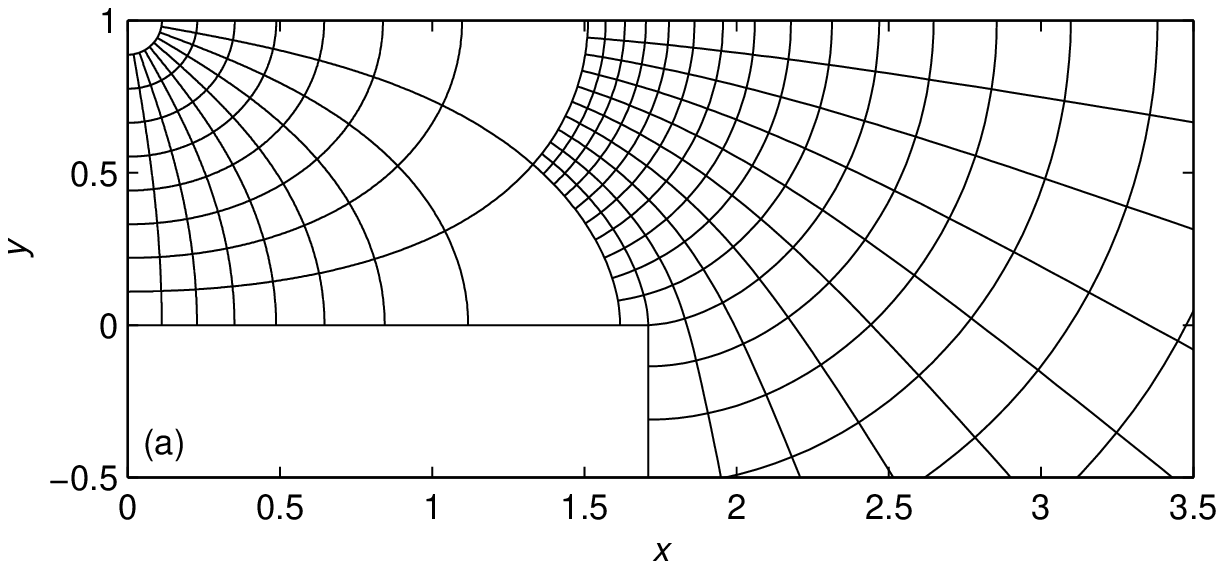}}%
\vspace{-4mm}
\par\vspace{5mm}
    \vspace{0mm}\mbox{\hspace{-1mm}\includegraphics
      [width=85mm,angle=0]{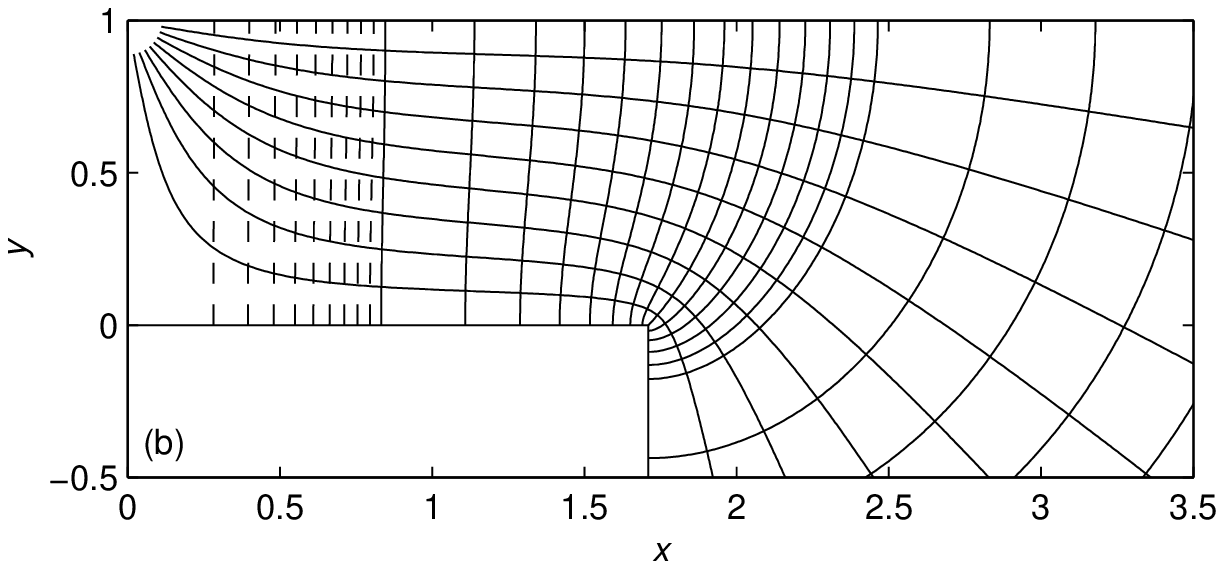}}%
\vspace{-4mm}
\par\vspace{5mm}
    \vspace{0mm}\mbox{\hspace{-1mm}\includegraphics
      [width=82mm,angle=0]{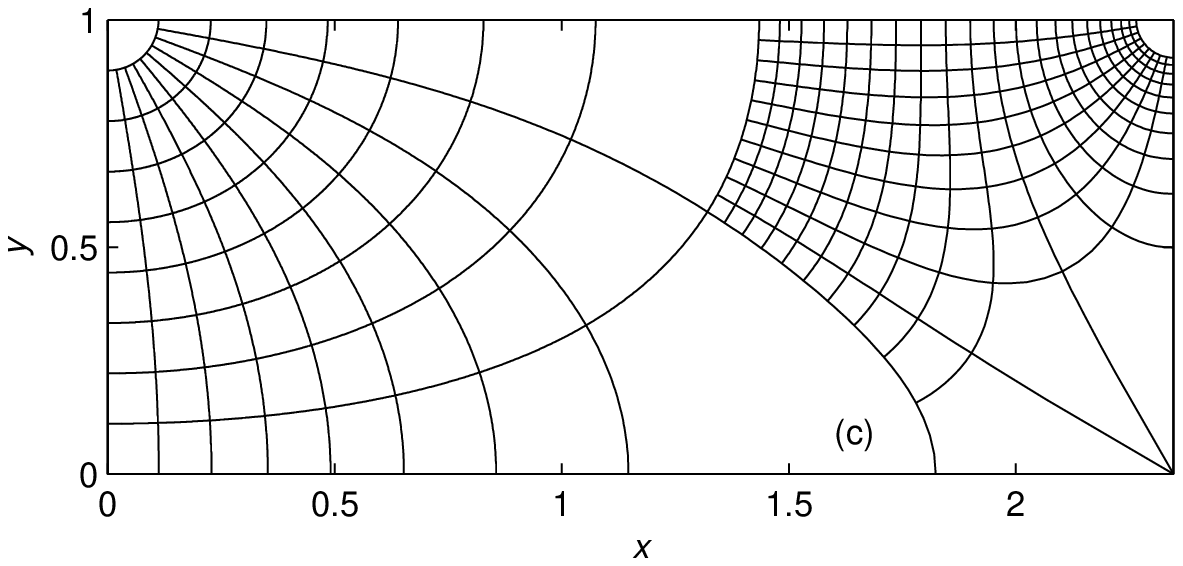}}%
      \vspace{-3mm}
\end{center}
\caption
{Extended transverse Mercator projection. (a)~shows the graticule and
(b)~the convergence and scale for Gauss--Kr\"uger projection.  (c)~shows
the graticule for the Thompson projection.  The ellipsoid parameters are
the same as for Fig.~\ref{graticule}.  The graticules in (a) and (c) are
the same as in Fig.~\ref{graticule}(b) with the addition of $1^\circ$
lines for $80^\circ < \lambda < 90^\circ$ and $-10^\circ \le \phi <
0^\circ$.  In (b), the lines emanating from the top left corner are
lines of constant meridian convergence, $\gamma$, at $10^\circ$
intervals.  The dog-legged line joining $(0,1)$, $(0,0)$, $(1.71,0)$,
and $(1.71,-\infty)$ represents $\gamma = 0^\circ$.  The line $y = 1$
gives $\gamma = 90^\circ$.  The other lines (running primarily
vertically in the figure) are lines of constant scale $k$.  The solid
lines show integer values of $k$ for $1 \le k \le 15$ and multiples of 5
for $15 \le k \le 35$.  The line segment joining $(0,0)$ and $(0,1)$
gives $k = 1$.  The dashed lines show lines of constant $k$ at intervals
of $0.1$ for $1 < k < 2$.}
\label{extend}
\end{figure}
It is possible to extend the mapping by moving to the ``right'' of the
equator in Fig.~\ref{graticule}(b).  If the complex phase of
$i\tilde\chi$ includes the interval $[0, \tfrac 3/2 \pi]$, an
``extended'' domain for the mapping may be defined by the union of
$0^\circ \le \phi \le 90^\circ$, $0^\circ \le \lambda \le 90^\circ$ and
$-90^\circ < \phi \le 0^\circ$, $\lambda_0 \le \lambda \le
90^\circ$.  The rule for analytic continuation is that the second region
is reached by a path from the central meridian which goes north of the
branch point.  This is equivalent to placing the cut so that it emanates
from the branch point in a south-westerly direction.  Following this
prescription, the range of the mapping now consists of the union of
$0 \le x$, $0 \le y/(k_0 a) \le E$ and $K' - E' \le x/(k_0 a)$, $y \le
0$.

Figures~\ref{extend}(a) and (b) illustrate the properties of the
Gauss--Kr\"uger projection in this extended domain.  These figures use an
ellipsoid with eccentricity $e = \frac1{10}$ (as in
Fig.~\ref{graticule}) and with $a = 1/E = 0.6382$ and $k_0 = 1$.  The
branch point then lies at $\phi = 0^\circ$, $\lambda = 81^\circ$ or $x =
(K' - E')/E \approx 1.71$, $y = 0$.  Symmetries can now be employed to
extend the mapping with arbitrary rules for how to circumvent the branch
point.  The symmetries are equivalent to placing mirrors on the four
lines segments: $0 \le x$, $y = 1$; $x = 0$, $0 \le y \le 1$; $0 \le
x \le 1.71$, $y = 0$; and $x = 1.71$, $y \le 0$.  Compare
Fig.~\ref{extend}(a) with \citet[Fig.~53(b)]{koenig51}.

Figure~\ref{extend}(c) shows the graticule of the Thompson projection in
the extended domain; the
range of this mapping is the rectangular region shown, $0 \le x/(k_0
a) \le K'$, $0 \le y/(k_0 a) \le K$.  The extended Thompson projection
has reflection symmetry on all the four sides of Fig.~\ref{extend}(c).
In transforming from Thompson to Gauss--Kr\"uger, the right angle at the
lower right corner of Fig.~\ref{extend}(c) expands by a factor of 3 to
$270^\circ$ to produce the outside corner at $x = 1.71$, $y = 0$ in
Fig.~\ref{extend}(a).  The top right corner of Fig.~\ref{extend}(c)
represents the south pole and this is transformed to infinity in the
extended Gauss--Kr\"uger projection.  Despite the apparent similarities,
the behavior of the extended Thompson projection near the north and
south poles (the top left and top right corners in Fig.~\ref{extend}(c))
is rather different.  Although the mapping at north pole is conformal,
the mapping at the south pole in the extended domain is not.  The
difference in longitude between the two meridians represent by the top
and right edges of Fig.~\ref{extend}(c) is $90^\circ e$ instead of
$90^\circ$.

My implementation of the exact mapping provides the option of using the
extended domain.  The round-off errors quoted in Sect.~\ref{exact}
($9\,\mathrm{nm}$ for double precision and $5\,\mathrm{pm}$ for extended
precision) apply to the extended domain for $\phi > -15^\circ$.  Beyond
this line, the errors grow because of the contraction of $w$ space near
the south pole (at $\phi = -58^\circ$, the error is about
$1\,\mathrm{mm}$).

\section{Conclusion}\label{conclusion}

The algorithms presented here allow the transverse Mercator projection
to be computed with an accuracy of a few nanometers.  Implementations of
these algorithms are included in GeographicLib \citep{geographiclib17}
which also provides (a)~the set of test data used to check the
implementations, (b)~Maxima code for the exact mapping (with
arbitrary precision), (c)~Maxima code for generating the Kr\"uger series
to arbitrary order, and (d)~the Kr\"uger series to 30th order.  The web
page
\url{http://geographiclib.sf.net/tm.html}
provides quick links to all these resources.

The work described in this paper made heavy use of the computer algebra
system \citet{maxima} both for carrying out the series expansions for
Kr\"uger's method and for generating the high accuracy test data.  The
latter is invaluable when developing complex algorithms with an accuracy
close to machine precision.  Other computer algebra systems offer
similar capabilities; but Maxima is one of the few that is free.

My emphasis in developing these algorithms was in their accuracy.
Nevertheless the resulting implementations are reasonably fast.  On a
$2.66\,\mathrm{GHz}$ Intel processor and compiled with g++, the time for
the mappings implemented with the 6th order series method is
$1.91\,\mathrm{\mu s}$; this is the combined time for a forward and a
reverse mapping including the computation of the convergence and scale
in each case.  This time is insensitive to number of terms retained in
the sum due to the efficiency of Clenshaw summation---changing this to 4
(resp.~8) decreases (resp.~increases) the time by only $1\%$.  Skipping
the calculation of the convergence and scale reduces the time by $15\%$.
Using a trigonometric series and Clenshaw summation for the conversions
between geographic and conformal latitude (as proposed
by \citet{engsager07}) decreases the time by $18\%$.  The exact
algorithms (which are accurate over the entire ellipsoid) are $5$--$6$
times slower.  The 6th order series method is comparable in speed to
Geotrans 3.0 even though the latter is much less accurate and does not
return the convergence and scale.

Here are some recommendations for users of the transverse Mercator
projection.  Do not use algorithms based on the formulas given
by \citet{thomas52}---they are unnecessarily inaccurate.  Instead use
the Kr\"uger series, truncating Eqs.~(\ref{radext}), (\ref{alphaext}),
and (\ref{betaext}) to order $n^6$.  With double precision, this gives
an accuracy of $5\,\mathrm{nm}$ for distances up to $3900\,\mathrm{km}$
from the central meridian.  If the mapping is needed at greater
distances from the central meridian, use the algorithm based on the
exact mapping (an accuracy of $9\,\mathrm{nm}$).  If greater accuracy is
needed, use extended precision with either method (extending the series
method to 8th order).  When implementing these algorithms, use the test
set to verify that the errors are comparable with those given here.

\section*{Acknowledgments}

I would like to thank Rod Deakin and Knud Poder for helpful discussions.

\bibliography{geod}

\begin{thebibliography}{22}
\providecommand{\natexlab}[1]{#1}
\providecommand{\url}[1]{\texttt{#1}}
\providecommand{\urlprefix}{URL }
\expandafter\ifx\csname urlstyle\endcsname\relax
  \providecommand{\doi}[1]{doi:\discretionary{}{}{}#1}\else
  \providecommand{\doi}{doi:\discretionary{}{}{}\begingroup
  \urlstyle{rm}\Url}\fi
\providecommand{\eprint}[2][]{\url{#2}}

\bibitem[{Bugayevskiy and Snyder(1995)}]{bugayevskiy95}
L.~M. Bugayevskiy and J.~P. Snyder, 1995, \emph{Map Projections: A Reference
  Manual} (Taylor \& Francis, London),
  \urlprefix\url{http://www.worldcat.org/oclc/31737484}.

\bibitem[{Bulirsch(1965)}]{bulirsch65}
R.~Bulirsch, 1965, \emph{Numerical calculation of elliptic integrals and
  elliptic functions}, Num. Math., \textbf{7}(1), 78--90,
  \doi{10.1007/BF01397975}.

\bibitem[{Carlson(1995)}]{carlson95}
B.~C. Carlson, 1995, \emph{Numerical computation of real or complex elliptic
  integrals}, Numerical Algorithms, \textbf{10}(1), 13--26,
  \doi{10.1007/BF02198293}, \eprint{math/9409227}.

\bibitem[{Clenshaw(1955)}]{clenshaw55}
C.~W. Clenshaw, 1955, \emph{A note on the summation of {C}hebyshev series},
  Math. Tables Aids Comput., \textbf{9}(51), 118--120,
  \urlprefix\url{http://www.jstor.org/stable/2002068}.

\bibitem[{Dozier(1980)}]{dozier80}
J.~Dozier, 1980, \emph{Improved algorithm for calculation of {UTM} and geodetic
  coordinates}, Technical Report NESS 81, NOAA,
  \urlprefix\url{http://fiesta.bren.ucsb.edu/~dozier/Pubs/DozierUTM1980.pdf}.

\bibitem[{Engsager and Poder(2007)}]{engsager07}
K.~E. Engsager and K.~Poder, 2007, \emph{A highly accurate world wide algorithm
  for the transverse {M}ercator mapping (almost)}, in \emph{Proc. XXIII Intl.
  Cartographic Conf. (ICC2007), Moscow}, p. 2.1.2.

\bibitem[{Geotrans(2010)}]{geotrans}
Geotrans, 2010, \emph{Geographic translator, version 3.0},
  \urlprefix\url{http://earth-info.nga.mil/GandG/geotrans/}.

\bibitem[{Hager \emph{et~al.}(1989)Hager, Behensky, and Drew}]{utmups}
J.~W. Hager, J.~F. Behensky, and B.~W. Drew, 1989, \emph{The universal grids:
  {U}niversal {T}ransverse {M}ercator ({UTM}) and {U}niversal {P}olar
  {S}tereographic ({UPS})}, Technical Report TM 8358.2, Defense Mapping Agency,
  \urlprefix\url{http://earth-info.nga.mil/GandG/publications/tm8358.2/TM8358_%
2.pdf}.

\bibitem[{Karney(2010)}]{geographiclib17}
C.~F.~F. Karney, 2010, \emph{Geographic{L}ib, version 1.7},
  \urlprefix\url{http://geographiclib.sf.net}.

\bibitem[{K\"onig and Weise(1951)}]{koenig51}
R.~K\"onig and K.~H. Weise, 1951, \emph{Mathematische {G}rundlagen der
  H\"oheren {G}eod\"asie und {K}artographie}, volume~1 (Springer, Berlin).

\bibitem[{Kr\"uger(1912)}]{krueger12}
J.~H.~L. Kr\"uger, 1912, \emph{Konforme {A}bbildung des {E}rdellipsoids in der
  {E}bene}, New Series~52, Royal Prussian Geodetic Institute, Potsdam,
  \doi{10.2312/GFZ.b103-krueger28}.

\bibitem[{Kuittinen \emph{et~al.}(2006)Kuittinen, Sarjakoski, Ollikainen,
  Poutanen, Nuuros, T\"atil\"a, Peltola, Ruotsalainen, and Ollikainen}]{jhs154}
R.~Kuittinen, T.~Sarjakoski, M.~Ollikainen, M.~Poutanen, R.~Nuuros,
  P.~T\"atil\"a, J.~Peltola, R.~Ruotsalainen, and M.~Ollikainen, 2006,
  \emph{{ETRS89}---j\"arjestelm\"a\"an liittyv\"at karttaprojektiot,
  tasokoordinaatistot ja karttalehtijako}, Technical Report JHS 154, Finnish
  Geodetic Institute, {A}ppendix 1, Projektiokaavart,
  \urlprefix\url{http://docs.jhs-suositukset.fi/jhs-suositukset/JHS154/JHS154_%
liite1.pdf}.

\bibitem[{Lagrange(1770)}]{lagrange70}
J.~L. Lagrange, 1770, \emph{Nouvelle m\'ethode pour r\'esoudre les
  \'equa\-tions litt\'erales par le moyen des s\'eries}, in
  \emph{Oeuvres}, volume~3, pp. 5--73 (Gauthier-Villars, Paris, 1869), reprint of
  M\'em. de l'Acad. Roy. des Sciences de Berlin {\bf 24}, 251--326,
  \urlprefix\url{http://books.google.com/books?id=YywPAAAAIAAJ&pg=PA5}.

\bibitem[{Lambert(1772)}]{lambert72}
J.~H. Lambert, 1772, \emph{Anmerkungen und {Z}us\"atze zur {E}ntwerfung der
  Land- und {H}immelscharten}, number~54 in Klassiker ex. {W}iss.
  (Engelmann, Leipzig, 1894), translated into English by W. R. Tobler as {\it Notes
  and Comments on the Composition of Terrestrial and Celestial Maps}, Univ. of
  Michigan (1972),
  \urlprefix\url{http://books.google.com/books?id=o_s_MR3NUD4C}.

\bibitem[{Lee(1976)}]{lee76}
L.~P. Lee, 1976, \emph{Conformal projections based on {J}acobian elliptic
  functions}, Cartographica, \textbf{13}(1, Monograph 16), 67--101,
  \doi{10.3138/X687-1574-4325-WM62}.

\bibitem[{Ludwig(1943)}]{ludwig43}
K.~Ludwig, 1943, \emph{Die der transversalen {M}ercatorkarte der {K}ugel
  entsprechende {A}bbildung des {R}otationsellipsoids}, J. Reine Angew. Math.,
  \textbf{185}(4), 193--230, \doi{10.1515/crll.1943.185.193},
  \urlprefix\url{http://resolver.sub.uni-goettingen.de/purl?GDZPPN002175576}.

\bibitem[{Maxima(2009)}]{maxima}
Maxima, 2009, \emph{A computer algebra system, version 5.20.1},
  \urlprefix\url{http://maxima.sf.net}.

\bibitem[{Olver \emph{et~al.}(2010)Olver, Lozier, Boisvert, and Clark}]{dlmf10}
F.~W.~J. Olver, D.~W. Lozier, R.~F. Boisvert, and C.~W. Clark, editors, 2010,
  \emph{{NIST} Handbook of Mathematical Functions} (Cambridge Univ. Press),
  \urlprefix\url{http://dlmf.nist.gov}.

\bibitem[{Stuifbergen(2009)}]{stuifbergen09}
N.~Stuifbergen, 2009, \emph{Wide zone transverse {M}ercator projection},
  Technical Report 262, Canadian Hydrographic Service,
  \urlprefix\url{http://www.dfo-mpo.gc.ca/Library/337182.pdf}.

\bibitem[{Thomas(1952)}]{thomas52}
P.~D. Thomas, 1952, \emph{Conformal projections in geodesy and cartography},
  Special Publication 251, U.S. Coast and Geodetic Survey,
  \urlprefix\url{http://docs.lib.noaa.gov/rescue/cgs_specpubs/QB275U35no251195%
2.pdf}.

\bibitem[{Wallis(1992)}]{wallis92}
D.~E. Wallis, 1992, \emph{Transverse {M}ercator projection via elliptic
  integrals}, Technical Report NPO-17996, JPL.

\bibitem[{Whittaker and Watson(1927)}]{whittaker27}
E.~T. Whittaker and G.~N. Watson, 1927, \emph{A Course of Modern Analysis}
  (Cambridge Univ. Press), 4th edition, reissued in Cambridge Math. Library
  Series (1996).

\end{thebibliography}
\end{document}